\documentstyle[twoside]{article}

\catcode`\@=11
\long\def\@makefntext#1{
\protect\noindent \hbox to 3.2pt {\hskip-.9pt  
$^{{\eightrm\@thefnmark}}$\hfil}#1\hfill}		

\def\@makefnmark{\hbox to 0pt{$^{\@thefnmark}$\hss}}	
	
\def\ps@myheadings{\let\@mkboth\@gobbletwo
\def\@oddhead{\hbox{}
\rightmark\hfil\eightrm\thepage}   
\def\@oddfoot{}\def\@evenhead{\eightrm\thepage\hfil
\leftmark\hbox{}}\def\@evenfoot{}
\def\sectionmark##1{}\def\subsectionmark##1{}}

\oddsidemargin=\evensidemargin
\newcounter{sectionc}\newcounter{subsectionc}\newcounter{subsubsectionc}
\renewcommand{\section}[1] {\vspace{12pt}\addtocounter{sectionc}{1} 
\setcounter{subsectionc}{0}\setcounter{subsubsectionc}{0}\noindent 
	{\tenbf\thesectionc. #1}\par\vspace{5pt}}
\renewcommand{\subsection}[1] {\vspace{12pt}\addtocounter{subsectionc}{1} 
	\setcounter{subsubsectionc}{0}\noindent 
	{\bf\thesectionc.\thesubsectionc. {\kern1pt \bfit #1}}\par\vspace{5pt}}
\renewcommand{\subsubsection}[1] {\vspace{12pt}\addtocounter{subsubsectionc}{1}
	\noindent{\tenrm\thesectionc.\thesubsectionc.\thesubsubsectionc.
	{\kern1pt \tenit #1}}\par\vspace{5pt}}
\newcommand{\nonumsection}[1] {\vspace{12pt}\noindent{\tenbf #1}
	\par\vspace{5pt}}

\newcommand{\textlineskip}{\baselineskip=13pt}
\newcommand{\smalllineskip}{\baselineskip=10pt}

\def\eightcirc{
\begin{picture}(0,0)
\put(4.4,1.8){\circle{6.5}}
\end{picture}}
\def\eightcopyright{\eightcirc\kern2.7pt\hbox{\eightrm c}} 

\newcommand{\copyrightheading}[1]
	{\vspace*{-2.5cm}\smalllineskip{\flushleft
        {\footnotesize Los Alamos archives: gr-qc/9411051 #1}\\
        {\footnotesize $\eightcopyright$\, H.C. Rosu, Nuovo Cimento B 108
        (1993) 313-329
         }\\
	 }}

\def\abstracts#1#2#3{{
	\centering{\begin{minipage}{4.5in}\baselineskip=10pt\footnotesize
	\parindent=0pt #1\par 
	\parindent=15pt #2\par
	\parindent=15pt #3
	\end{minipage}}\par}} 

\renewenvironment{thebibliography}[1]
	{\frenchspacing
	 \ninerm\baselineskip=11pt
	 \begin{list}{\arabic{enumi}.}
        {\usecounter{enumi}\setlength{\parsep}{0pt}     
	 \setlength{\leftmargin 12.7pt}{\rightmargin 0pt} 
         \setlength{\itemsep}{0pt} \settowidth
	{\labelwidth}{#1.}\sloppy}}{\end{list}}

\newcounter{itemlistc}
\newcounter{romanlistc}
\newcounter{alphlistc}
\newcounter{arabiclistc}

\def\@citex[#1]#2{\if@filesw\immediate\write\@auxout
	{\string\citation{#2}}\fi
\def\@citea{}\@cite{\@for\@citeb:=#2\do
	{\@citea\def\@citea{,}\@ifundefined
	{b@\@citeb}{{\bf ?}\@warning
	{Citation `\@citeb' on page \thepage \space undefined}}
	{\csname b@\@citeb\endcsname}}}{#1}}

\newif\if@cghi
\def\cite{\@cghitrue\@ifnextchar [{\@tempswatrue
	\@citex}{\@tempswafalse\@citex[]}}
\def\citelow{\@cghifalse\@ifnextchar [{\@tempswatrue
	\@citex}{\@tempswafalse\@citex[]}}
\def\@cite#1#2{{$\null^{#1}$\if@tempswa\typeout
	{IJCGA warning: optional citation argument 
	ignored: `#2'} \fi}}

\def\@refcitex[#1]#2{\if@filesw\immediate\write\@auxout
	{\string\citation{#2}}\fi
\def\@citea{}\@refcite{\@for\@citeb:=#2\do
	{\@citea\def\@citea{, }\@ifundefined
	{b@\@citeb}{{\bf ?}\@warning
	{Citation `\@citeb' on page \thepage \space undefined}}
	\hbox{\csname b@\@citeb\endcsname}}}{#1}}

\def\@refcite#1#2{{#1\if@tempswa\typeout
        {IJCGA warning: optional citation argument
	ignored: `#2'} \fi}}

\def\refcite{\@ifnextchar[{\@tempswatrue
	\@refcitex}{\@tempswafalse\@refcitex[]}}

\def\pmb#1{\setbox0=\hbox{#1}
	\kern-.025em\copy0\kern-\wd0
	\kern.05em\copy0\kern-\wd0
	\kern-.025em\raise.0433em\box0}

\def\fnt#1#2{\footnotetext{\kern-.3em
	{$^{\mbox{\scriptsize #1}}$}{#2}}}

\font\tenrm=cmr10
\font\tenit=cmti10 
\font\tenbf=cmbx10
\font\bfit=cmbxti10 at 10pt
\font\ninerm=cmr9

\font\eightrm=cmr8

\def\qed{\hbox{${\vcenter{\vbox{			
   \hrule height 0.4pt\hbox{\vrule width 0.4pt height 6pt
   \kern5pt\vrule width 0.4pt}\hrule height 0.4pt}}}$}}

\begin{document}



\normalsize\textlineskip
\thispagestyle{empty}
\setcounter{page}{1}

\copyrightheading{}			

\vspace*{0.88truein}

\centerline{\bf HYBRIDIZING THE SKYRMION WITH AN ANTI-DE SITTER BAG}
\vspace*{0.035truein}
\vspace*{0.37truein}
\centerline{\footnotesize HARET C. ROSU}
\vspace*{0.015truein}
\centerline{\footnotesize\it Instituto de F\'{\i}sica,
Universidad de Guanajuato, Apdo Postal E-143, Le\'on, Gto, Mexico}
\centerline{\footnotesize\it rosu@ifug3.ugto.mx}
\baselineskip=10pt
\vspace*{10pt}
\vspace*{0.225truein}

\vspace*{0.21truein}
\abstracts{
I discuss a phenomenological model of the nucleon in which a
small anti-de Sitter bag is placed into the Skyrmion configuration.
Such a bag has a timelike boundary and allows naturally the
Cheshire Cat Principle. Very important in this model is the membrane
of the bag, the 3-dimensional time-space manifold $S^{1}\times S^{2}$,
in which topological techniques will come into play.
}{}{}


\textlineskip                  
\vspace*{12pt}                 

\vspace*{1pt}\textlineskip	
\vspace*{-0.5pt}
\noindent


\noindent




\noindent


\vskip 2cm

\noindent
Contents 
\noindent\\
1. Introduction 
\noindent\\
2. Review of de Sitter and anti-de Sitter spaces
\noindent\\
3. AdS femtocosmology 
\noindent\\
4. Cheshire Cat Principle and the AdS bag
\noindent\\
5. The pion and the AdS bag
\noindent\\
6. Conclusions and perspectives
\noindent\\
References

\vskip 2cm
\noindent
This is a slightly revised version of IC-26/92 and of Nuovo Cimento
B 108, 313-329 (1993). IC-26/92 has been written in January 1992, and revised
in October 1992 and November 1994. The companion paper is Nuovo Cimento
A 107, 1543-1547 (1994) [gr-qc/9406017(IFUG-17b/94)].

\vskip 1cm

\noindent
IFUG-17a/94 [$\cal H \cal C \cal R $]

\newpage
\section{ Introduction}

Since 1932 the atomic nuclei are considered to be made of protons
and neutrons. This is due mainly to the fact that their degree
of virtuality is very low (8 MeV as compared to 939 MeV).
Nevertheless quantum chromodynamics (QCD) \cite{kn:0},
 arising so powerfully
in the seventies to explain hadronic structure, forced us to look at
hadrons and atomic nuclei as merely low energy phenomenology of that
 theory.

A decisive step forward on such a route was the MIT bag model, which
 subsequently gave birth to an impressive number of other bags. Most
 remarkably, a model of the fifties due to Skyrme became prominent
 recently \cite{kn:1a}. As it is well known the attractive
  features of the
 Skyrmion are connected with some fresh mathematics streaming
 into the realm of hadron and nuclear physics from such fields
 like topology and differential geometry. At the same time, Skyrmions
 have been proven to have a good matching to Yukawa theory as well as
 to bag models through hybridization. Thus, Skyrmions may be looked
 upon as a necessary development in low energy hadronic and nuclear
 phenomenology.
 In the past a number of important results were established within
 Skyrmion physics, some of which are :

 i) Skyrme model is an effective low energy/large $N_{c}$ limit of
 QCD \cite{kn:1}.

ii) The baryon number has a topological nature and belongs to the
homotopy group $\pi_{3}(S^{3})= Z$ \cite{kn:2}. The winding baryon
number of the mappings $S^{3}\rightarrow S^{3}$ is
$$B=\int d^{3}x B_{0}~,   \eqno(1.1)$$
where $B_{0}$ is the time component of the conserved topological current
$$B_{\mu}=\frac{1}{24\pi ^{2}}\epsilon_{\mu\nu\alpha\beta}
trL^{\nu}L^{\alpha}L^{\beta}~.  \eqno(1.2)$$
The $L$ matrices, $L_{\mu}=U^{\dagger}\partial _{\mu} U$,
 are left topological currents of $SU(2)$ matrices
parametrizing the pion fields
$$U(x)=\exp [2i\tau^{a}\phi _{a}/F_{\pi}]~,  \eqno(1.3)$$
where $\tau ^{a}$ are the Pauli matrices, $\phi _{a}$ are the pion
 fields, and $F_{\pi}$ is the pion decay constant used to normalize
the pion fields.

iii) The Wess-Zumino term added to the Skyrme Lagrangian fuses two
unwanted discrete symmetries making the model viable \cite{kn:3}.

The Skyrme Lagrangian is invariant separately under
$$ P_{1}\;: \vec{x}\rightarrow -\vec{x}  \eqno(1.4)$$
and
$$P_{2}\;: U\rightarrow U^{\dagger},   \eqno(1.5)$$
which are combined into the parity transformation $P=P_{1}\cdot P_{2}$.
However, processes like
$K^{+}K^{-}\rightarrow \pi ^{+}\pi ^{-}\pi ^{0}$ break $P_{1}$ and
 $P_{2}$. Therefore an action term is required, which breaks these
  symmetries keeping the combined parity $P$ invariant. The
   Wess-Zumino-Witten (WZW) term is acting like that for the SU(3)
 Skyrmions. It is a 5-dimensional integral over a disk $D$ whose
 boundary is the four dimensional space-time
 $$S_{WZW}=-\frac{iN_{c}}{240\pi ^{2}}
 \int_{D} d^{5}x\epsilon^{\alpha\beta\gamma\delta\epsilon}TrL_{\alpha}
 L_{\beta}L_{\gamma}L_{\delta}L_{\epsilon}~.   \eqno(1.6)$$
 The coefficient in front of the integral, proportional to $N_{c}$,
 is determined by computing electromagnetic processes such as
 $\pi ^{0}\rightarrow 2\gamma$, for this action, with an additional
 gauge-invariant coupling to the electromagnetic field, and comparing
 it with the direct QCD calculation \cite{kn:3a}.

iv) The Cheshire Cat Principle: if a bag is introduced into the
Skyrme soliton as a kind of defect \cite{kn:4}, then the dynamics
is independent of the boundary conditions at the bag surface. This
is an exact statement in two dimensions only \cite{kn:5}.

Usually, one will make use of a chiral coupling between quarks and pions
at the bag surface, i.e.,
$U_{5}= \exp (i\vec{\tau}\cdot \vec{\phi} \gamma_{5}/f_{\pi})$.

In the following we want to introduce into the Skyrmion a rather special
bag possessing anti -de Sitter (AdS) geometry and to make an analysis of the
 consequences of proceeding in this way. The first two sections of the
 paper are short surveys  on the topics of de Sitter (dS) spaces and
 strong gravity (femtocosmology). They were written for the convenience
of the readers
 (who are supposed to belong mostly to the Skyrmion society). The third
 section shows that the AdS bag is very appealing for the
 Cheshire Cat Principle. In the fourth section we comment on the pion
 as the AdS Goldstone boson. Finally, we add some conclusions. The
 present paper is an updated and improved version of a talk delivered
 at Poiana Brasov School in the summer of 1988 \cite{kn:5a}.

\section{ Review of de Sitter and anti-de Sitter spaces}

de Sitter spaces are ``respectable" objects in theoretical physics
ever since 1917. There is an extensive literature on various aspects
concerning these spacetimes. Very good introductory material could be
found in a number of well known books and review papers \cite{kn:6}.

de Sitter universes are maximally symmetric solutions of Einstein
cosmological equations
$$R_{\mu \nu}-\frac{1}{2} Rg_{\mu \nu} + \Lambda g_{\mu \nu}=
\kappa T_{\mu \nu}            \eqno(2.1)~,$$
with no energy-momentum tensor, $T_{\mu \nu}=0$. Thus, de Sitter
 solutions are vacuum ones, and it was precisely this feature as
 related to a variant of Mach principle which was very much discussed
 in 1917, when W. de Sitter introduced his universe in Physics \cite{kn:6a}.

On the other hand, spaces of constant negative curvature were brought
into Physics by A.A. Friedman in his second famous cosmological
paper of 1924 written at the suggestion of his good friend Tamarkin.
Mathematically, these spaces had already at that time a history of
one hundred years \cite{kn:6b}, starting with Lobachevsky and Bolyai,
 although
Minding was the first one who studied surfaces of constant negative
curvature (1839), and of course we must remember Beltrami who
established the connection between hyperbolic geometry and these
surfaces in 1868.

de Sitter space was much investigated in cosmology, as it was the first
cosmological model possessing a red-shift (i.e., a direct observable).
After 1980, many authors have dealt with this space in connection with
the inflationary hypothesis asserting that a dS phase dominated the
very early Universe ($t\leq 10^{-35}$ s; $T\geq 10^{15}$ GeV).

AdS space, on the other hand, turned out to be very important in
extended supergravity, in Kaluza-Klein theories, and more generally in
microcosmos. This is
due to a number of reasons, amongst which positive definiteness of the
energy operator, the causal relationship one could set up in the AdS
case, the possibility to construct an S-matrix and also the analogue
of Penrose transform to a generalized twistor space.

Considered as Riemannian spaces, dS universes being of constant
curvature have the curvature tensor given by the following formula
$$R_{\mu \nu \rho \sigma}= a^{2}(g_{\mu \rho} g_{\nu \sigma} -
g_{\mu \sigma} g_{\nu \rho})~.
\eqno(2.2)$$
Locally, dS spaces are expressed by the equation
$$R_{\mu \nu}-1/4Rg_{\mu \nu}=0   \eqno(2.3)$$
In mathematics such spaces are called Einstein spaces \cite{kn:6c}.
If, as physicists, we prefer to keep a cosmological constant in our line of
reasoning, then we have the following well-known relation
 $$\Lambda=1/4 R~.   \eqno(2.4)$$
 Assuming spherical symmetry, de Sitter line element turns into the
 form first written by de Sitter as follows
 $$ds^{2}=(1-1/3\Lambda r^{2})dt^{2}-(1-1/3 \Lambda r^{2})^{-1}dr^{2}
 -r^{2}(d\theta^{2}+\sin^{2}\theta d\phi^{2})~.    \eqno(2.5)$$
 The embedding in a 5-dimensional pseudoeuclidean space is given
 in the case $\Lambda>0$ by
 $$x_{0}=(\frac{3}{\Lambda})^{\frac{1}{2}}(1-\frac{1}{3}\Lambda r^{2})^
 {\frac{1}{2}}\sinh [(\frac{\Lambda}{3})^{\frac{1}{2}}t]$$
 $$x_{1}=(\frac{3}{\Lambda})^{\frac{1}{2}}(1-\frac{1}{3}
 \Lambda r^{2})^{\frac{1}{2}}\cosh [(\frac{\Lambda}{3})^{\frac{1}{2}}t]
  \eqno(2.6)$$
 $$x_{2}=r\sin \theta \cos \phi$$
 $$x_{3}=r\sin\theta \sin \phi$$
 $$x_{4}=r\cos \theta~.$$
 These embedding equations turn the dS universe into the hypersurface
 $$+x_{0}^{2}-x_{1}^{2}-x_{2}^{2}-x_{3}^{2}-x_{4}^{2}= -\frac{3}{\Lambda}~.
  \eqno(2.7)$$

For the AdS space ($\Lambda<0$) the embedding formulas are
$$x_{0}=(\frac{3}{-\Lambda})^{\frac{1}{2}}(1-\frac{1}{3}\Lambda r^{2})
^{\frac{1}{2}}\sin [(\frac{-\Lambda}{3})^{\frac{1}{2}}t]$$
$$x_{1}=(\frac{3}{-\Lambda})^{\frac{1}{2}}(1-\frac{1}{3}\Lambda r^{2})
^{\frac{1}{2}}\cos [(\frac{-\Lambda}{3})^{\frac{1}{2}}t] \eqno(2.8)$$
$$x_{2}=r\sinh \theta \cos \phi$$
$$x_{3}=r\sinh \theta \sin \phi$$
$$x_{4}=r\cosh \theta~.$$
The pseudoeuclidean coordinates are constrained this time to the
hypersurface
$$+x_{0}^{2}-x_{1}^{2}-x_{2}^{2}-x_{3}^{2}+x_{4}^{2}=-\frac{3}{\Lambda}~.
   \eqno(2.9)$$
The four-dimensional quadrics (2.7) and (2.9) are a four-hyperboloid
with one sheet in the dS case and a four-hyperboloid with two sheets in
the AdS case. Their topology is $R\times S^{3}$ and $S\times R^{3}$,
respectively.

The line element (2.5) could also be written in the Robertson-Walker
(expansion) form
$$ds^{2}=d\tau^{2}-\frac{3}{\Lambda}\cosh^{2}[(\frac{\Lambda}{3})^{
\frac{1}{2}}\tau ](d\chi^{2}+\sin^{2}\chi d\Omega^{2}),\; \; {\rm for}
\Lambda >0   \eqno(2.10)$$
$$ds^{2}=d\tau^{2}+\frac{3}{\Lambda}\cos^{2}[(\frac{-\Lambda}{3})^
{\frac{1}{2}}\tau](d\chi^{2} + \sinh^{2} \chi d\Omega^{2}),\; \; {\rm for}
\Lambda<0~. \eqno(2.11)$$
The above expressions make it obvious that dS universe is closed, i.e.,
with a spherical 3-dimensional geometry, while the AdS universe is open, i.e.,
with a hyperbolic 3-dimensional geometry.

For the AdS case, it is worth noting that the scale factor (the radius
 of the universe in the older literature) is an oscillating function,
 with a maximum amplitude
 $$r_{max}=\left(\frac{3}{-\Lambda}\right)^{\frac{1}{2}}~.  \eqno(2.12)$$
The last useful form of the line element is the conformal one
$$ds^{2}=(a\cos\rho)^{-2}[dt^{2}-d\rho^{2}-\sin^{2}\rho (d\theta^{2}
+\sin^{2}\theta d\phi^{2})]~,    \eqno(2.13)$$
where the radial variable has the range $0<\rho<\pi /2$, the other
ranges being the common ones. The parameter $a$ is given by
 $R=-12a^{2}$.

As usual when looking to spacetimes as 4-dimensional geometries, we
have to consider first of all their kinematical group of isometries.
The de Sitter and Anti-de Sitter groups are the most simple and direct
generalization of the Poincare group, which can be obtained from the
first ones by In\"{o}n\"{u}-Wigner contraction procedure.

In homogeneous coordinates, the Absolute (i.e., the invariant quadric) of
the Lorentz group is given by
$$+\xi _{1}^{2}+\xi _{2}^{2}+\xi _{3}^{2}-c^{2}\xi _{4}^{2}=0~,
 \eqno(2.14)$$
 whereas the Absolutes of dS/AdS groups are determined by
$$+\xi _{1}^{2}+\xi _{2}^{2}+\xi _{3}^{2}-c^{2}\xi _{4}^{2}\pm R^{2}
\xi _{5}^{2}=0~.  \eqno(2.15)$$

One can see immediatelly that this generalization of the Lorentz group
 is equivalent to the minimal introduction of cosmology through a
  constant curvature of positive value for dS space and of negative
  value for AdS space.

We shall use mostly geometrical units: c=1;$\Lambda=3$;$\hbar=1$.

The quadratic form (2.9) corresponding to the AdS space has to be
invariant under the action of the isometry group SO(3,2). The ten
Killing vectors are of the usual ``angular momentum" type
$$K_{AB}=-(x_{A}\partial_{B}-x_{B}\partial_{A})  \eqno(2.16)$$
and satisfy the Lie algebra
$$[K_{AB},K_{CD}]=+K_{AC}\eta _{BD}+K_{BC}\eta _{AD}-K_{AD}\eta _{BC}
-K_{BD}\eta _{AC}~,     \eqno(2.17)$$
where $\eta_{AB}=diag(+1,-1,-1,-1,+1)$.
When one is passing to the quantum theory the covariant generators have to
be used
 $$\hat{M}_{AB}=-iK_{AB}~.     \eqno(2.18)$$
 To be more specific one has noncommutative translations
 $$[\hat{P}_{\alpha},\hat{P}_{\beta}]=R\hat{M}_{AB}~.   \eqno(2.19)$$
 If we denote
 $$\hat{M}_{A4}=r\hat{P}_{a}~,  \eqno(2.20)$$
 where r is the AdS radius of curvature, and
 $$\hat{M}_{AB}=\hat{M}_{\alpha \beta}  \eqno(2.21)$$
 for the other generators, we get the compact Lie product (commutator)
 of the type (2.17). In\"{o}n\"{u}-Wigner contraction simply means
 $r\rightarrow 0$.

Since SO(3,2) is a noncompact group, the hermiticity of the operators,
($\hat{M}_{AB}=\hat{M}_{AB}^{+}$),implies the unitarity of only the
 infinite dimensional representations. The compact operators are
 $\hat{M}_{04}$ and $\hat{M}_{ab}$ and the noncompact ones are
 $\hat{M}_{0a}$ and $\hat{M}_{a4}$. The corresponding physical
  quantities are obtained from the contracted Lie algebra (the
  Poincar\'{e} one). Since under contraction $\hat{M}_{04}$ goes
  into $\hat{P}_{0}$, which is the time translation operator,
$\hat{M}_{04}$ could be identified with the energy operator; also
$\hat{M}_{ab}$are angular momentum operators; the noncompact operators
$\hat{M}_{0a}$ and $\hat{M}_{a4}$ correspond to the momentum operators
and Lorentz boost operators, respectively. The center of SO(3,2) is
bidimensional. The two Casimir operators are the following
$$\hat{C}_{2}=1/2 \hat{M}^{AB} \hat{M}_{AB}$$
$$\hat{C}_{4}=1/2\hat{M}^{AB} \hat{M}_{BC} \hat{M}^{CD} \hat{M}_{DE}~.
  \eqno(2.22)$$

The Casimir operators of SO(3,2) are similar to the Casimir operators
of the Poincar\'{e} group. In the representations of the
Poincar\'{e} group, the latter ones have the eigenvalues $m^{2}$
and $m^{2}s(s+1)$, where m denotes the mass and s the spin. However
the SO(3,2) representations are not usually labeled by means of
$\hat{C}_{2}$ and $\hat{C}_{4}$ eigenvalues. Instead of this,$E_{0}$,
the smallest value of the AdS energy (in units of a) and s, the
minimal eigenvalue of the total angular momentum are used (they
correspond to the maximal compact subgroup of SO(3,2) which is
$SO(3,2)\otimes SO(3,2)$). The reason for that is to be found in the
fact that the Casimir eigenvalues do not uniquely specify the
representations of noncompact groups. Another general fact for such
groups is the occurrence of indecomposable representations
(noncompletely reducible), of much help in doing generalized dS
quantum electrodynamics in parallel to the usual Minkowski QED \cite{kn:7}.

The eigenvalue of the Casimir operator $\hat{C}_{2}$ could be expressed
in terms of $E_{0}$ and s as follows
$$C_{2}=E_{0}(E_{0}-3)+s(s+1)~.    \eqno(2.23)$$
Having two timelike directions the group O(3,2) must contain as a
factor group the conformal group of 3 dimensions, i.e., the group of
symmetries of a cone in a 3-dimensional pseudo-Euclidian space with
signature $(+,-,-)$. Consider the 5-cone $\eta_{AB} x^{A}x^{B}=0)$ and
define homogeneous coordinates in the following way
$$y^{\mu}=x^{\mu}/(x_{3}+x_{4})~,   \eqno(2.24)$$
where $\mu =0,1,2$.
The linear action of O(3,2) on the 5-cone given by
$$x_{A}^{'}=L_{A}^{B} x_{B}~,    \eqno(2.25)$$
with $L_{A}^{B} \in O(3,2)$ induces by (2.24) a nonlinear action on
the Minkowski space of $y_{\mu}$ denoted as Mink(y). The action of
O(3,2) on Mink(y) is not effective, i.e., we have two identity
transformations on Mink(y) coming from O(3,2), namely the subgroup
$Z_{2}=\{\Lambda \in O(3,2): \Lambda x=\pm x\}$. The conformal group is
 defined as: $C(1,2)\sim O(3,2)/Z_{2}$. The group O(3,2) has four
 components. That one connected to the identity is
 $$SO_{0}(3,2)=\{L\in O(3,2): {\rm det} L=1; L_{0}^{0}L_{4}^{4}-L_{0}^{4}
 L_{4}^{0} >1\}~.     \eqno(2.26)$$
 The other components are obtained through combined reversals of the
 signs of det L and of the ``time order". One can see that
 $SO(3,2)\subset C(1,2)$. To pass from the component connected to the
 identity to some other component, discrete operations are required
 $${\cal T}:=\{x^{0}\rightarrow -x^{0}; x^{4}\rightarrow x^{4};x^{i}
 \rightarrow x^{i}\}    \eqno(2.27)$$
$${\cal P}:=\{x^{0}\rightarrow x^{0}; x^{4}\rightarrow x^{4}; x^{i}
\rightarrow -x^{i}\}    \eqno(2.28)$$
$${\cal A}:=\{x^{0}\rightarrow -x^{0}; x^{4}\rightarrow -x^{4}; x^{i}
\rightarrow -x^{i}\}~.   \eqno(2.29)$$
$\cal T$,$\cal P$, and $\cal A$ stand for time reversal, space parity,
 and antipodal inversion (or parity), respectively. $\cal A$ commutes
 with all elements of SO(3,2) and has the involutory eigenvalues $\pm 1$.
 $\cal P$ and $\cal A$ are related by
 $${\cal P}= {\cal V}{\cal A} {\cal V}^{-1}~,   \eqno(2.30)$$
where ${\cal V}\in SO_{0}(3,2)$ is given by
 $${\cal V} :=\{x^{0}\rightarrow -x^{0}; x^{4}\rightarrow -x^{4}; x^{i}
 \rightarrow x^{i}\}~.   \eqno(2.31)$$

For the spin-1/2 fields in dS space we have to add a ``spin part" to the
SO(3,2) ``angular part"
$$\hat{J}_{AB}=\hat{M}_{AB}+\hat{S}_{AB}~.       \eqno(2.32)$$
The charge conjugation matrix of AdS space is the standard $4\times 4$
matrix of Minkowski space \cite{kn:8}. For more sophisticated
applications in supergravity and KK theory where AdS manifolds
occur as natural backgrounds, it is worth noting that the SUSY
 extension of SO(3,2), which is the supergroup Osp(4/N) has 4N Killing
 spinors playing the same role in the spinor space as do Killing vectors
 in the boson space. In the case N=1, all linear unitary irreducible
 representations of AdS SUSY have been classified by Heidenreich \cite{kn:9}.
The best (super)review has been written in 1986 \cite{kn:10}.
The most complete discussion of the $SO_{0}(3,2)$
  representations has been provided by Angelopoulos \cite{kn:11}.
  For more physics-oriented treatments one should see Gibbons \cite{kn:8},
as well as Nicolai \cite{kn:12}.

The representations $D(E_{0},s)$ are unitary iff
$$E_{0}\geq s+1, \; \; s=1, 3/2,2,5/2, ...$$
$$E_{0}\geq s+1/2, \; \; s=0,1/2~.      \eqno(2.33)$$
Fields with $s\geq 1$ reach the unitarity limit when $E_{0}=s+1$ and
since they contract to the massless discrete helicity representations
of the Poincar\'e group, they are called ``massless". Fields with s=0,
1/2 have $E_{0}=s+1/2$ in the unitary limit but have no Poincar\'e
limit. They are not massless and have been called singletons by Dirac
\cite{kn:13}, who discovered them. As a tribute, Fronsdal \cite{kn:14},
dubbed  the boson singleton as the RAC (i.e., D(1/2,0)),
and the fermion one as the DI, (i.e., D(1,1/2)). They have been under
the focus of many authors, especially in their superform since they
could be related to super p-branes \cite{kn:14a}.
 From the group-theoretical standpoint (super)singletons are
included in an ultrashort (super)multiplet.

Singletons have two remarkable properties namely, the one-singleton
 states are unobservable and the two-singleton states are always
  massless, i.e.,
  $$Singleton \otimes Singleton= \oplus Massless \; Reps~.   \eqno(2.34)$$
  The essential point to stress is that the quantum field theory of
  singletons is intimately related to the spatial infinity of AdS space
  which is the famous manifold $S^{1}\times S^{2}$. Geometrically it is
  precisely this AdS boundary that give birth to singleton fields. The
  fact that the AdS boundary is the remarkable 3-dimensional
  $S^{1}\times S^{2}$ manifold makes us foresee many topological
  methods entering the field. The very interesting and well-developed
topology (in Mathematics) of this ``Membrane at the End of the Universe"
\cite{kn:15}, is at the same time the topology of
the ``Nucleon Membrane", as I would call it.

We have to recall also the strong analogy between quarks and singletons
which has been put forth by Flato and Fronsdal \cite{kn:16}, and as a
matter of fact, one can look at them as completely equivalent objects.
This is, in our opinion, the most important property of AdS space still
to be fully exploited \cite{kn:16a}.

Many encumbrances reveal themselves when someone is going to tackle the
problem of quantizing fields in AdS background. This is due to the
$S^{1}\times R^{3}$ topology of AdS. Avis, Isham, and Storey
\cite{kn:17} have discussed in a very clear way the quantization of
scalar fields in AdS space. The things to be emphasized are first,
the impossibility to specify Cauchy data on a single spatial
 hypersurface, in other words, AdS is a non-global hyperbolic manifold.
 Second, it possesses closed timelike curves. This is considered as an
 unwanted property and is avoided by going to AdS universal covering
 space (CAdS) which is a global manifold with a $R\times S^{3}$
 topology allowing to be conformally mapped onto half of the Einstein
 static universe (ESU). The boundary at spatial infinity remains
  timelike even for the CAdS case implying loss or gain of information
  through spatial infinity in finite coordinate time. Such a situation
  is unacceptable for a well defined quantum field theory. In order to
  develop a consistent quantization scheme Avis, Isham, and Storey
  have introduced reflective boundary conditions (RBC) at spatial
  infinity, in other words, they have put there reflecting walls.
  We shall use again results from  \cite{kn:17} in section 3.

Until about five years ago, AdS quantum field theory has been a flourishing
topic, especially due to SUSY, SUGRA,and KK theories \cite{kn:18}, but
meanwhile the activity in the field has diminished as the result of other
fashionable discoveries.
We recall here the parallel and less ambitious use of dS groups and
their SUSY extensions as dynamical (spectrum generating)
groups \cite{kn:19}. In this case
one is not interested so much in the structure of the ``constituents",
but rather in their relative motions and for that purpose the
 oscillator representations are recommendable. AdS space may be
  considered as a general relativistic version of the harmonic
 oscillator potential with the curvature playing the role of the
 harmonic oscillator coupling constant. In this way, dynamical spectra
(e.g., Regge trajectories) are reproducible by means of AdS dynamical
group. On the other hand, dS groups have also been used as
gauge groups \cite{kn:20}, and more
practically in some powerful techniques for separation of variables
in partial differential equations \cite{kn:21}.

\section{ AdS femtocosmology}

As has been put by Allen and Turyn, ``de Sitter spaces
 interest different people
 for different reasons" \cite{kn:22}. Our goal is to make a connection
 between AdS physics and Skyrmion physics. This could be done by
 inserting into the Skyrmion configuration a bag (or a defect in the
 terminology of G. Brown) possessing AdS geometry. It is a kind of
 (femto)cosmology, which is known to have some tradition \cite{kn:23}.

An AdS bag was beautifully discussed for the first time by Salam and
Strathdee in the f-g theory context \cite{kn:24}. Among precursors,
we name Prasad, who developed somewhat similar ideas before the Bag
Era \cite{kn:25}. Subsequently, a dS bag was compactly described by
Haba \cite{kn:26}. Romaniuk studied an AdS bag with ``dust" as the
gravitational source \cite{kn:27}. Important developments on the AdS
geometrical confinement problem have been worked out by the Nijmegen
 group \cite{kn:28}.

Femtocosmology could be considered also as a variant of the bimetric
general relativity \cite{kn:29}, applied at the hadron spacetime scale
where the coexistence of the two metrics is supposed to occur, the
common Minkowski one and the AdS one.

The basic idea in femtocosmology of considering hadrons as strongly
curved femtouniverses is accomplished by scaling down the AdS Universe
by giving to the curvature scalar a value fixed rather arbitrarily by
some hadronic scale, or equivalently by introducing a huge cosmological
constant
$$\Lambda_{h}\sim 10^{26} {\rm cm}^{-2}\sim10^{-2} {\rm GeV}^{2}~.
\eqno(3.1)$$
It is worth noting that such a huge gravitational effect of Higgs
origin is nevertheless considered small on the scale of quantum
 gravity. We are still at several orders of magnitude away from the
 Planck scale.

 What is the reason making possible the scaling
up and/or down of universes? In other terms, what does make the
curvature scalar arbitrary? The answer to these important questions
is given by the hyperbolic geometry. Let us take the hyperbolic
surface ${\cal H}^{2}$, though the discussion holds good for any number of
dimensions. It is a mathematical fact that the angle of parallelism in
${\cal H}^{2} $ is given by the well-known relation
$$\tan[1/2 \Pi(x)]={\rm e}^{-x/\kappa}~,        \eqno(3.2)$$
where $\kappa$ is the hyperbolic curvature, which one could express
 more intuitively in terms of the distance $ \cal D$ corresponding to an
  angle of parallelism of $\pi /4$ radians. The connection is
 $$ \kappa=0.2 \cal D~.      \eqno(3.3)$$
 The point is that hyperbolic geometry asserts the completely arbitrary
 character of the distance $\cal D$. It follows that Mathematics allows
 the scaling of universes of constant curvature. In AdS spaces the
 radius of curvature appears on the same level as the velocity of
 light, as one can see from (2.15). Hence it may be considered as a
 constant of physics very similar to a fundamental length \cite{kn:29a}.
 There is
 no etalon for the curvature, and consequently it is difficult to
 consider it as a physical constant. Its nature is purely geometrical.
 However the spontaneous breaking process is the common procedure by
 which geometrical constants are introduced into physics. Since AdS
 group is a subgroup of the conformal group C(4,2), the spontaneous
 breaking of the conformal symmetry leaves only SO(3,2) symmetry in the
 real world. This symmetry breaking was studied in a very
  straightforward way by Fubini \cite{kn:30}. On the other hand,
  de Sitter spaces are in turn conformally flat and therefore a rescaled
  conformally coupled field does not recognize as special the length
  scale set by the curvature of the spacetime in which it lives up to
  the conformal anomaly.

In order to fix the existing value $r\sim 1$ fm, we shall proceed rather
 arbitrarily
and use a grand-unified Higgs field. Indeed it is essential to note that
one can get the AdS geometry by means of a scalar field with a huge mass,
since in this case the scalar energy-momentum tensor simulates the vacuum
one, more exactly in the limit $(\dot{\varphi} /\varphi )^{2}\ll M^{2}$
 we get the vacuum equation of state $p=-\epsilon$.

The following procedure is well known in relation to the spontaneous
symmetry breaking rules \cite{kn:31}. The Lagrangian of a complex
scalar field is written down
$${\cal L} =\partial_{\mu} \varphi^{+} \partial^{\mu} \varphi +
\mu^{2} \varphi^{+} \varphi -\lambda (\varphi^{+} \varphi )^{2}
 \eqno(3.4)$$
 The field $\varphi =\varphi_{1}+i\varphi_{2}$ has hermitic components.
One of them has a nonzero vacuum expectation value denoted by $\sigma$.
Then the fields are shifted by $\sigma$. The new Lagrangian in the
shifted fields $\phi_{1}$ and $\phi_{2}$ will be
$$L=\hat {L} +1/2 \mu^{2} \sigma^{2}-1/4 \lambda \sigma^{4}~.
 \eqno(3.5)$$
 All the operator terms have been included in $\hat {L}$.
 Stability arguments give
 $$ \sigma^{2}=\mu^{2}/\lambda~.       \eqno(3,6)$$
 The vacuum expectation value of the Hamiltonian will be
 $$<H>_{vac}=-\mu^{2} \sigma^{2}/4= <T_{\mu \nu}>_{vac}~.   \eqno(3.7)$$
 Now, one could use Einstein equations to get
 $$<T_{\mu \nu}>_{vac} =\Lambda/\kappa~.     \eqno(3.8)$$
 From (3.4) we can get a negative cosmological constant of the form
 $$\Lambda =-1/4 \mu^{2} \sigma^{2}~.    \eqno(3.9)$$
 The maximal value of the AdS radius will follow as
 $${\cal R} =(12/\mu^{2} \sigma^{2})^{1/2}~.    \eqno(3.10)$$
 In order to have an AdS vacuum at the hadron scale we need a massive
 GUT Higgs field with $\mu \sim 10^{15} GeV$. The Higgs pressure is
 given by the quantity $\Lambda/\kappa$.

\section{ Cheshire Cat Principle and the AdS bag}

Throughout this section we shall stay in two dimensions. Cheshire Cat
Principle (CCP) is a rigorous result in this case and has to do with
the boundary conditions at the bag surface. CCP has still to be
generalized to four dimensions. We recall that quantum field theory in
two dimensions is dominated by the remarkable complete equivalence of
fermion and boson theories \cite{kn:32}. CCP which has been put forth
in 1985 is just a consequence of this general feature \cite{kn:5}.
For fermion -boson equivalence in four dimensions see the appendix in
the paper of Rubakov \cite{kn:32a}.

Our physical scenary is that of a skyrmion with an AdS bag inside it.
The 2-dimensional AdS space (AdS$_{2}$) has the following metric in
intrinsic coordinates $t, \rho$
$$ds^{2}=(a\cos \rho)^{-2}[dt^{2}-d\rho^{2}]~,    \eqno(4.1)$$
where $-\pi/2<\rho<\pi/2 $ and $-\pi \leq t \leq \pi$ in AdS, the latter
range extending to the whole real axis for CAdS. The image of
AdS$_{2}$ in Minkowski space is a finite spatial interval whose boundary
 consists of 2 points, $\rho=\pm \pi/2$.

The CCP concept has been introduced and developed at Nordita \cite{kn:5};
see also \cite{kn:33}. We shall discuss only the abelian case, i.e.,
fermions of only one colour and flavor. CCP is opposite to confinement
allowing fermions to leak out of the bag (in the membrane region in our
model) and finally in the skyrmion region they show up as a soliton of
the boson representation. It is some highly nontrivial topological
dynamics in the $S^{1}\times S^{2}$ membrane which is involved in the
CCP process. The idea of the standard CCP is to make the Euler-Lagrange
equations of the total action to coincide at least partially with the
bosonization relations. This is achieved if one adds to the fermioic
and bosonic action a boundary term of the form \cite{kn:34}
$$ S_{B}=\int_{B} d\epsilon^{\mu}(1/2 n_{\mu} \overline{\psi}
 e^{i\sqrt{4\pi} \gamma_{5} \phi} \psi)~,    \eqno(4.2)$$
where $d\epsilon^{\mu} =d\epsilon \cdot n^{\mu}$ is a surface
area element, $n^{\mu}$ being the bag normal. Setting the total surface
variation equal to zero, one can get the following two boundary equations
 $$i\not n \psi =n^{2} e^{i\sqrt{4\pi \gamma _{5}}\; \phi} \psi$$
 $$n\partial \phi =\sqrt{\pi} \; \overline{\psi} \not n \gamma_{5} \psi~.
  \eqno(4.3)$$
  These are slightly modified Coleman relations, or the $\psi$-equation
  and the $\phi$-equation. The latter one is just the normal component
of the axial current bosonization relation
$$\partial ^{\mu} \phi =\sqrt{\pi}\; \overline{\psi} \gamma_{\mu}
\gamma_{5} \psi~.   \eqno(4.4)$$
As known, in the two-dimensional world the axial current can be
expressed easily through the vector current.

The tangential component of the bosonization relation is nicely presented
in the ``criminal story" of Nielsen and Wirzba \cite{kn:35}. CCP is
brought in while one is trying to derive the $\psi$-equation. In order
not to get a decoupled $\psi$-equation, one should take care to allow
the vanishing of the determinant of a $2\times 2$ matrix, whose entries
are derived by imposing locality, renormalizability and hermiticity
conditions. For details see the work of Nadkarni and Nielsen \cite{kn:34}. It
happens that one finds solutions only for spacelike bag
normals, i.e., timelike bag walls. This important constraint is
automatically satisfied if the bag has AdS geometry, because of the
timelike surface at spatial infinity ($\rho =\pi/2$) of AdS or CAdS.
In their seminal paper Avis, Isham and Storey have introduced reflective
boundary conditions (RBC), later used extensively in AdS SUSY, but also
``transparent" boundary conditions (TBC), which in our opinion are
perfectly suited for discussing CCP. Imposing proper boundary conditions
is essential for a good quantization scheme in AdS space. The
difficulties with quantization in AdS background are related to the
fundamental periodicities of AdS. As AdS is not globally hyperbolic, one
has to pass to another conformally flat space, the Einstein static
universe (ESU), proceed with quantization there, and only then mapping
 back to AdS. The detailed analysis \cite{kn:17} shows that Cauchy
data must be specified on two spacelike surfaces separated in time by
$\pi$. The periodicities of the mode functions in AdS and ESU imply that
only the following sum of ``conserved" quantities is really constant
$$P_{a} =Q_{a} (\tau) + Q_{a} (\tau + \pi)~,    \eqno(4.5)$$
where
$$Q_{a}(\tau )=\int_{\tau ={\rm const}}
T_{0\nu} \xi_{a}^{\nu} g^{00} \sqrt{-g}d\rho~.    \eqno(4.6)$$
The symbols of energy-momentum tensor, Killing vectors and metric tensor
are the standard ones.

The unusual conservation law (4.5) dictates the recirculation of the
energy, angular momentum and conserved conformal quantities lost through
the timelike interface at spatial infinity. In two dimensions the AdS
background has SO(2,1) invariance and the conserved quantities correspond
to the ``translation" generator, dilatation, and restricted spatial
conformal transformation, that is to the three conformal Killing vectors
of the background geometry.

In curved space bosonization rules have to be modified in order to
compensate the scale change of fermion masses by a position-dependent
interection with which the bosons have to be provided. Thus for massive
fermions the $\phi$-equation turns into
$$n\partial \phi =\sqrt{\pi}\; \Omega \;\overline{\psi}\not n \gamma_{5}
\psi~,     \eqno(4.7)$$
where the conformal factor $\Omega =(\cos \rho )^{-1}$ for AdS.

A related problem is that of Dirac operators with space-dependent
masses \cite{kn:366}.

It is worth noting that the helicity eigenstates in AdS are linear
 combinations of antipodal eigenstates \cite{kn:36}. The antipodal
 reflection is a spatial reflection plus a translation in time by
 $\Delta t =\pi /a$. For a more intuitive analogy, the antipodal
 reflection is similar to the total reflection in which the
  electromagnetic field is leaking in the propagation-``forbidden"
medium causing a time ``delay". In this way one will not encounter
with AdS bags a helicity paradox as for MIT bags ( a boundary changes
always the helicity) since ``AdS-quarks" at the ``walls" suffer antipodal
reflections, a very appealing general fact.

\section{ The pion and the AdS bag}

Theoretically, the pion has many faces \cite{kn:37}. In QCD, the pion
may be considered as a single bound pair of valence quark and antiquark.
 Since u and d quarks are almost massless on hadronic scales, the
resulting chiral $SU(2)_{L} \times SU(2)_{R}$ symmetry of QCD suggests
the identification of the pion with the associated Goldstone boson. It
can be viewed also as a collective $q\overline{q}$ mode built on the QCD
vacuum \cite{kn:37}.

The most attractive model remains that of Nambu$\; \&$ Jona-Lasinio
 (NJL).
In this well-known model, the chiral symmetry is dynamically broken
through an effective chiral interection which moves the mass of the
pionic $q\overline{q}$ mode down to zero according to the Goldstone
theorem.

Many years ago, Castell has considered the problem of the Goldstone
boson in AdS space \cite{kn:38}. He also introduced an invariant
partial causal order possible only for AdS and not for dS. This time
order is given for $s^{2} \leq 0$ events by the Castell function
$${\cal C} (x,x^{'})={\rm signum}(x_{0}x_{4}^{'}-x_{4}x_{0}^{'})~,
 \eqno(5.1)$$
with the order values $\pm 1$ for ``before" and ``after".
The Laplace-Beltrami equation has been written by Castell in AdS
horospherical coordinates and he made use of a scheme of the Goldstone
theorem worked out by Bludman and Guralnik. The result of Castell is
that the Goldstone AdS boson is pseudoscalar and has a mass
$$ m_{G,C}= m_{\pi} = 3a~.     \eqno(5.2)$$
Castell's AdS pion (the representation D(3,0)) is the Goldstone boson of
the spontaneously broken conformal symmetry
 $SO(4,2) \rightarrow SO(3,2)$. The idea of connecting the pion to the
breaking of conformal symmetry was reinforced by Isham, Salam and
Strathdee \cite{kn:39} in an effective Lagrangian formalism in which
 the symmetry breaking term is taken to be in an irreducible
representation of the combined conformal and chiral groups. In such a
model the bare masses of the pion and dilaton field (introduced to
make the Lagrangian conformal invariant) are connected.

Since in the Skyrme model the baryon charge has topological origin
we have be extremely careful with the details of the Goldstone
theorem. According to Shamir and Park \cite{kn:39a},
 topological symmetries are never broken spontaneously whenever a local
 operator is used as the order parameter. Evading this by means of a
 delocalized operator, it can be shown that the ``symmetry breaking"
 condition reflects the effect of the non-local order parameter on
the boundary conditions at spatial infinity. A substantial part of the
norm of the order parameter is to be found at spatial infinity.

 The effects of a constant spacetime curvature on
Nambu-Goldstone bosons were studied by Inami and Ooguri \cite{kn:40},
who showed that even in AdS$_{2}$ a continuous symmetry may be broken
because the gravitational potential of this space, which confines
particles within a finite spatial interval, is acting also as an infrared
cutoff. The fluctuations of the Nambu-Goldstone mode decay exponentially
 near the spatial infinity. Also Verbin \cite{kn:41} has dealt with
the spontaneous symmetry breaking in AdS background of an O(2) scalar
doublet and obtained a massive Goldstone mode too.

\section{ Conclusions and perspectives}

The ultimate goal of Skyrmion physics is the explanation of atomic
nucleus on a much more sophisticated basis than the conventional
nuclear physics \cite{kn:kleb}, a commitment from the side of Mathematics.

One of the best candidates to achieve this difficult task is a hybrid
skyrmion. However, in our opinion, it is mandatory that the bag inside
the Skymion to be of AdS type and thus allow cosmology to enter the
field. With the AdS bag the requirements of of CCP are trivial and
many other results are foreseeable. Surely the boundary of AdS bag,
spreading all over an $S^{1}\times S^{2}$ membrane, will be of great
importance in the future Skyrmion physics. It is still to be decided
 if this membrane is leptodermous or pachidermous. It is well
known that totally antisymmetric tensor fields as well as gravitational
fields are able to produce closed membranes by quantum processes, such
 as instanton tunneling \cite{kn:41a}. The interesting fact of the
 process is that it could stop automatically as soon as the effective
 cosmological constant becomes less or equal to zero. I believe this
``neutralization of the cosmological constant by membrane creation"
 \cite{kn:41a} to be essential for further considerations on the lines
 of the present paper.

One may argue that because the spatial sections of AdS bubbles are
infinite in extent these are not the right bags to put into the
skyrmions. This objection is easy to reject. Heckman and Sch\"{u}cking
have shown in 1959 that the space ${\cal H }^{3}$ could be substituted
by a compact hyperbolic manifold iff some points are properly
 identified \cite{kn:42}. The resulting hyperbolic compact spaces
 have no continuous group of isometries as stated by a theorem of
 Yano and B\"{o}chner \cite{kn:43}. Killing vectors could be found at every
 point of these spaces but it is only a local property and no global
 Killing vector does exist. This is not an essential difficulty in
 femtocosmology. One should keep in mind also that in doing quantum field
theory on such spaces confined by means of topological constraints,
zero point energy phenomena (Casimir energy) are extremely important
\cite{kn:43aa}.

One expects the bags to become more relevant in relativistic and
ultrarelativistic nucleus-nucleus collisions, even though they are
by far more appropriate for static conditions \cite{kn:43a}.
 Some hints for the bags
to be AdS spaces come from the hadronic multiplicity distributions
where data are explained by parton branching probabilities \cite{kn:44}.
These probabilities have hypergeometric functions in their
integral kernels and very similar hypergeometric functions enter the
expressions for all sorts of field propagators in dS spaces \cite{kn:45}.
Moreover, a connection between QCD and femtocosmology
is suggested by some works of Calzetta et al. \cite{kn:46}, in which
they showed that the asymptotic freedom could be obtained by means of
high curvature.

Even a torsion field may be accomodated. P. Minkowski \cite{kn:47}
has shown how little the Einstein equations are modified if one introduces
a contorsion field of the form
$$K_{abc}=\epsilon_{abc} K  \; \; (a,b,c,=1,2,3)~,   \eqno(6.1)$$
where
$$K=\eta /R    \eqno(6.2)$$
with $\eta$ an arbitrary constant and $R$ the scale factor of the universe.
The only effect of such a field is to change the curvature index
parameter k (k=1,0,-1 for spherical, flat and hyperbolic universes
respectively) to the new value
$$k_{1}=k-\eta^{2}   \eqno(6.3)$$
A hyperbolic universe remains hyperbolic in the presence of a contorsion
field of the type (6.3). This is of course an interesting result.
Indeed, we could obtain hyperbolic femto-universes from a spherical
universe by locally creating contorsion fields with the contorsion index
$\eta =\pm \sqrt{2}$. Starting with a flat universe, we need the local
 contorsion index to be $\pm 1$.

Some other useful references are recorded at the end of the published
ones \cite{kn:more}.

\nonumsection{ Acknowledgments}
\noindent
The author would like to thank Professor Abdus Salam, the International
Atomic Energy Agency and UNESCO for hospitality at the International
Centre for Theoretical Physics, Trieste.

\nonumsection{References}

\end{document}